\documentclass[a4paper]{article}

\usepackage{icrc2013}

\title{Accounting for the Sun and the Moon in {\em Fermi}-LAT Analysis}

\shorttitle{Solar System Tools}

\authors{
Gudlaugur Johannesson$^{1}$,
Elena Orlando$^{2}$,
for the {\em Fermi}-LAT Collaboration.
}

\afiliations{
$^1$ Science Institute, University of Iceland, Reykjavik, IS-107, Iceland \\
$^2$ W. W. Hansen Experimental Physics Laboratory, Kavli Institute for Particle Astrophysics and Cosmology, Stanford University, Stanford, CA 94305 \\
}

\email{gudlaugu@glast2.stanford.edu; eorlando@stanford.edu}

\abstract{The Sun and the Moon are quiescent gamma-ray sources that are
clearly detectable in {\em Fermi}-LAT data. While moving through the sky, the Sun
and the Moon can be a significant background in the analysis of {\em Fermi}-LAT
data if they pass through the region of interest. Accurate modeling of their
intensity is needed in this case, accounting for the correct exposure of their
track along the sky. We present the Solar System Tools (SST) which calculate
the templates of the Sun's and the Moon's intensity in the sky for a given
observing period and a model of their steady emission. These tools are
available in the standard {\em Fermi}-LAT Science Tools.}


\begin{document}
\maketitle

%Begin a section.

\section{Introduction}
The Sun and the Moon are both quiet gamma-ray sources
\cite{SolarPaper,LunarPaper}.  They are so far the only known bright emitters of
$\gamma$-rays with fast celestial motion.  Their emission can be a significant
background in the analysis of sources near the ecliptic and it even
contributes over the entire sky \cite{EGBPaper}. Accurate modeling of the
emission from the Sun and the Moon is therefore needed for analysis of
$\gamma$-ray data.  This is especially true for variability studies of sources
near the ecliptic, where the emission from the Sun and the Moon can mimick a periodic signal.

In this proceedings we will describe the Solar System Tools (SSTs) which are a
set of tools designed to incorporate Solar and Lunar emission into {\em Fermi}-LAT
analysis.

\section{The Basic Idea}

To first order the emission from the Sun and the Moon can be considered
spherically symmetric \cite{SolarPaper,LunarPaper}.  It is therefore sufficient to model their emission as a
function of energy and angle from their center.  This is a trivial task if one
works in either Sun centered or Moon centered coordinates.  In those
coordinates, however, all the other sources are moving on the sky, making their
analysis tricky.  It is therefore essential to know the exposure as a function
of energy and angle from the center of the Sun and the Moon.  This is a non-trivial task because of
the movement of the Sun and the Moon and the {\em Fermi}-LAT effective area dependency
on the incident angle of the incoming photon \cite{LATPaper}.

To facilitate the calculation of exposure in
{\em Fermi}-LAT analysis the Science-Tools have what is called a livetime cube.  It is
a sky-map where
each pixel contains a histogram of the livetime of the instrument as a function
of instrument angle.  This information can then be used to quickly evaluate the
exposure given a set of instrument response functions (IRFs).  We have extended
this method to also bin the livetime as a function of distance from the moving
Sun or the Moon.  This allows for accurate evaluation of the exposure in Solar
or Lunar coordinates over the entire sky.  Convolving that with a model of the
Sun and the Moon allows us to create accurate prediction for their emission.

\section{Solar System Tools}
The SSTs are designed to produce a template of the Solar and Lunar emission,
given a model for their emission and an observing period.  These templates are
unique for each observing period and also depend on the cuts applied on the
data, especially the cut on zenith angle.  The SSTs produces the template as a
fits CCUBE which can be then incorporated in standard likelihood analysis with
\texttt{gtlike} in a similar manner as the diffuse emission.

The tools do not depend on a specific Lunar and Solar emission models but we do
provide provide models taken from \cite{LunarPaper} and \cite{SolarPaper}
respectively. The Solar inverse Compton (IC) emission model is calculated with the
stellarics\footnote{publicly available at
\url{http://sourceforge.net/projects/stellarics}} software \cite{PubSostice} and it is
in agreement with the model 1 in \cite{SolarPaper}.  This software can be used
to generate alternative models for the Solar IC emission, e.g. for different
electron spectra or modulation.
The FITS files of the models are:
\begin{itemize}
\item \texttt{solar\_profile\_v2r0.fits} (for the Sun, disk+IC)
\item \texttt{lunar\_profile\_v2r0.fits} (for the Moon)
\end{itemize}
They are available in the SST package.

The SST package contains 4 tools that will be described in detail in the
subsection below:
\begin{itemize}
\item \texttt{gtltcubesun}: This tool calculates the livetime cube binned in
both instrument angle and the angle from the center of either the Sun or the
Moon.
\item \texttt{gtltsumsun}: This tool sums up livetime cubes calculated with
\texttt{gtltcubesun}.  Due to the design of the tools it is advised to split the
livetime cube calculations into smaller time bins and sum up in the end.  It
also facilitates parallel execution of \texttt{gtltcubesun}.
\item \texttt{gtexphpsun}: Calculates the exposure as a function of energy,
position on the sky, and angle from the Sun or the Moon.  It uses the livetime
cube caluclated by \texttt{gtltcubesun} or \texttt{gtltsumsun}
\item \texttt{gtsuntemp}: Calculates the expected intensity on the sky as a
function of energy to be used in \texttt{gtlike} analysis.  It uses the exposure
map from \texttt{gtexphpsun}, a model of the emission from either the Sun or the
Moon as appropriate, and an exposure map from the science tool
\texttt{gtexpcube2}.
\end{itemize}

\subsection{gtltcubesun}
\begin{figure}
\centering
\includegraphics[width=0.5\textwidth]{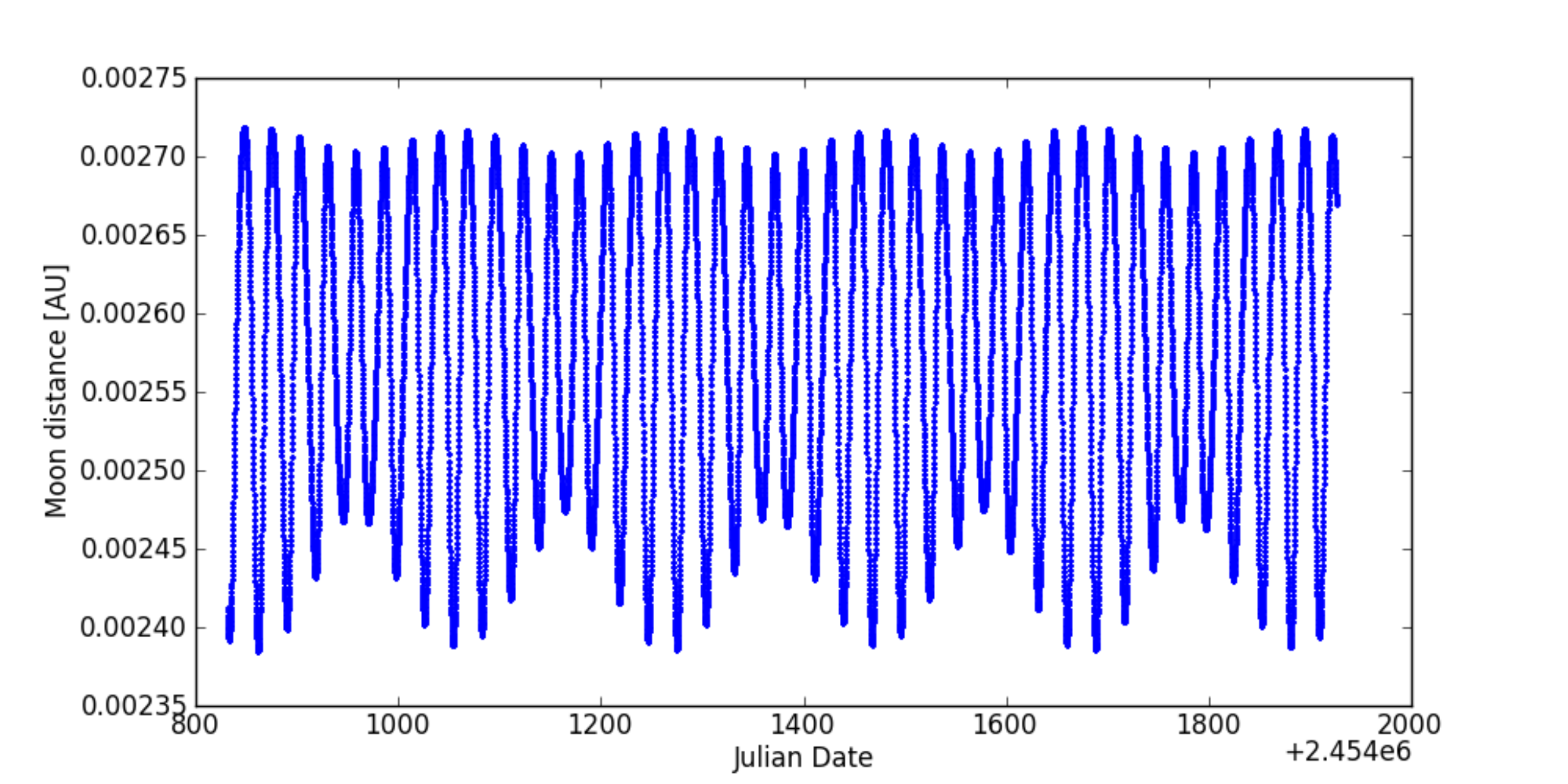}
\caption{The distance between the Earth and the Moon over the entire lifetime of
the {\em Fermi} mission.}
\label{fig:LunarOrbit}
\end{figure}

This tool calculates the integrated livetime binned in instrument coordinates
and the distance from either the Sun or the Moon.  It is an extended version of
the science tool \texttt{gtltcube}.  We use the astro library to calculate the
position of the Sun and the Moon on the sky.  The ellipticity of the Earth's and
Moon's orbits are taken into account by scaling the livetime with $1/d^2$, where
$d$ is distance to the moving body.  This correction is only applied for the
innermost $2.5$ degrees, corresponding to the disk emission.
Figure~\ref{fig:LunarOrbit} shows the distance to the Moon as a function of time
over the lifetime of the {\em Fermi} mission.  The change in distance causes
variation of about 25\% in the emission from the Moon.  While the Moon's orbit
does recess, the recession period is nearly 9 years so this effect does not
smooth out quickly.  The eccentricity of the Earth's orbit is much less than
that of the Moon, but this effect still causes around 5\% variation in the
emission from the Solar disk.

In addition to the parameters of the science tool \texttt{gtltcube} this tool
has the following parameters:
\begin{itemize}
\item {\em thetasunmax}: The maximum distance from the moving source used in
binning.  It should be 180 degrees for the Sun, but 0.5 degrees should be plenty for the
Moon.  Smaller values speed up the calculation and reduce the file size.
\item {\em powerbinsun}: The binning in angle from the moving source is done
evenly in $\cos(\alpha)^{1/p}$, where $\alpha$ is the angle from the moving
source and $p$ is this parameter.  We recommend 2.7 for the Sun and 2 for the
Moon.  Larger values result in smaller livetime cubes but less accuracy.
\item {\em body}: Specify the source, either SUN or MOON.
\end{itemize}

It is possible to modify the {\em binsz} parameter from the default value of
0.5 degrees but we do not recommend that.  Larger binning will cause a loss in
livetime because the binning in angle from the moving source is fixed at 0.25
degrees.  We have found that a value of 0.25 does not change the template
significantly for most analysis.

\subsection{ gtltsumsun}
This tool sums up livetime cubes generated by \texttt{gtltcubesun}.  It does some basic
tests to make sure the livetime cubes are compatible while adding them.  Due to
the way the storage is packed when calculating the livetime cubes it is
considerably faster to split the calculations of the livetime cube into smaller
time bins and sum them up in the end.  We recommend using one week time bins for
the Sun and month time bins for the Moon.  It is easy to split it up by using
the parameters {\em tmin} and {\em tmax} in \texttt{gtltcubesun}.

\subsection{gtexphpsun}
This tool calculates the exposure for different energies as a function of
distance from the Sun or Moon using the livetime cube generated by
\texttt{gtltcubesun} or \texttt{gtltsumsun}.
The tool has similar functionality as \texttt{gtexpcube2} but it supports only
HEALPix binning \cite{Gorski:2005}.  The energy binning parameters are identical to the ones from
\texttt{gtexpcube2}.  There is a single spatial binning parameter {\em binsz}
that specifies the approximate bin size to use in the HEALPix binning.  It
defaults to 0 where it uses the binning in the livetime cube.

\subsection{gtsuntemp}
This tool creates an intensity map appropriate for use in \texttt{gtlike}
analysis for the given observing period used to create the livetime cube.  It
requires a model profile for the emission of the moving source, the binned
exposure generated by \texttt{gtexphpsun}, and an exposure map generated by
\texttt{gtexpcube2}.  It is vital that the energy binning of the output
template and the input exposure maps are identical.  
The only model profile format currently supported is a
FITS file with three table extensions named: ANGLES, ENERGIES, SST$\_$PROFILE.
The first extensions specifies the angles used in the profile in degrees, the second
extension lists the energies used in the profile in MeV, and the third contains
the actual profile.  The profile is stored in a vector column, where each row
contains the profile as a function of angle for a specific energy.  The units of
the profile should be cm$^{-2}$sr$^{-1}$s$^{-1}$MeV${^-1}$. 

The tool works by first
calculating the expected photons sr-1 MeV-1 from the moving source using the
angular binned exposure map and the input profile. This is then turned into
average intensity by dividing with the normal exposure map. If the energy
binning of the Solar profile does not match that of the output map it uses
power-law interpolation.

\section{Example Templates}

\begin{figure*}[!t]
\centering
\includegraphics[width=\textwidth]{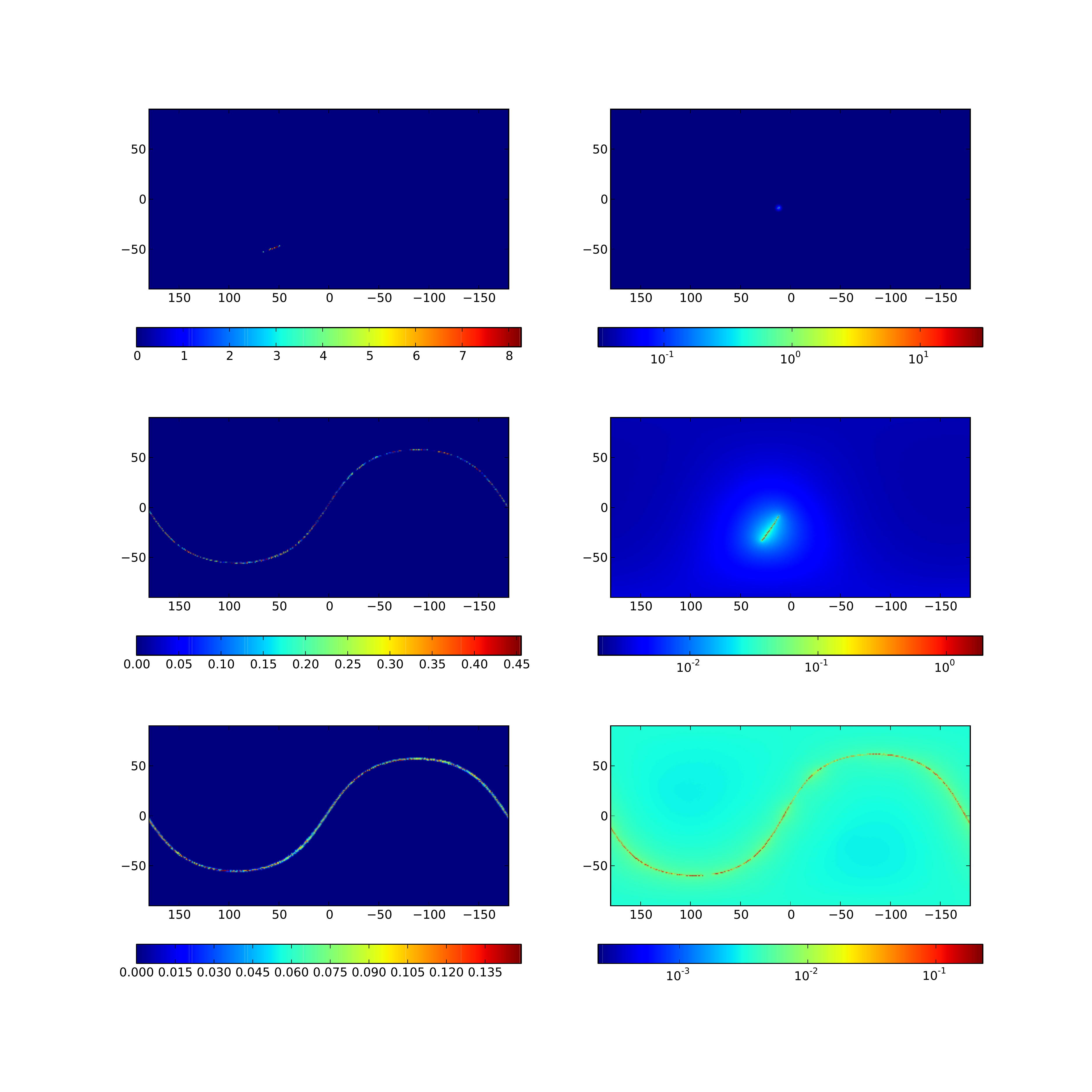}
\caption{Example templates using real observing profiles.  Left is the moon,
right is the sun. From top to bottom: 1 day, 1 month, 12 months. The templates
are full-sky in Galactic coordinates.  To better show the IC component of the
Sun, the Solar templates are shown on a logarithmic scale.  This is a template
for 100 MeV and the units are $10^{-6}/($cm$^{-2}$s$^{-1}$sr$^{-1}$MeV${-1})$. }
\label{fig:examples}
\end{figure*}

For illustration we have calculated a few templates using the SST package for
both the Sun and the Moon.  Figure~\ref{fig:examples} shows templates for the
Sun and the Moon for three different observing periods, a day, a month and a
year.  These templates were created using real observing profiles and the
starting period for all of them is January 1st 2009.  The templates are
not very sensitive to the IRFs used for creating them and these were created using
P7CLEAN\_V6 IRFs.

\section {Caveats}
The SST package assumes that the emission from the Sun and the Moon is constant in
time.  It therefore cannot account for rapidly varying emission such as Solar
flares.  It also does not account for variations in the emissivity due to
changes in CR density due to varying Solar activity.  To account for time
variations in the emission one needs to split the calculations into time periods
where the emission can be considered constant.  Then one can create an average
intensity map by adding the intensity maps multiplied by the exposure for the
smaller periods and in the end divide with the exposure for the entire period.
This is in fact the method employed by \texttt{gtltsuntemp} so it should be
accurate, as long as the assumption of a fixed emission within each smaller
period holds.  Note that this method should not be used to account for Solar
flares in the background.  It is much safer to exclude the time period around
the Solar flare in the analysis.

\vspace{0.5cm}
\footnotesize{{\bf Acknowledgment:}{ E.O. acknowledges
NASA grant \mbox{NNX12AO73G}. The {\em Fermi}-LAT Collaboration acknowledges
support from a number of agencies and institutes for both development and the
operation of the LAT as well as scientific data analysis. These include NASA
and DOE in the United States, CEA/Irfu and IN2P3/CNRS in France, ASI and INFN
in Italy, MEXT, KEK, and JAXA in Japan, and the K.~A.~Wallenberg Foundation,
the Swedish Research Council and the National Space Board in Sweden.
Additional support from INAF in Italy and CNES in France for science analysis
during the operations phase is also gratefully acknowledged.}}

\end{document}